\documentclass[aps,showpacs,pre,twocolumn]{revtex4}

\usepackage{psfig}

\newcommand{\Eqn}[1]{Eq.~(\ref{#1})}     
\newcommand{\Sec}[1]{Section~\ref{#1}}     
     
\newcommand{\Fig}[1]{Fig.~\ref{#1}}
\newcommand{\ve}{v}
\newcommand{\nlb}{SST75,DR76,SC77,CC82}

\begin{document}
\title{Transport on percolation clusters with power-law distributed bond 
strengths: when do blobs matter?}
\author{Mikko Alava}
\affiliation{Helsinki University of Technology, Laboratory of Physics, 
P.O.Box 1100, FIN-02015 HUT, Finland}
\affiliation{NORDITA, Blegdamsvej 17, DK-2100 Copenhagen, Denmark} 
\author{Cristian F.~Moukarzel} 
\affiliation{CINVESTAV M\'erida, F\'{\i}sica Aplicada, 97310 M\'erida
  Yucat\'an,   M\'exico} \date{\today}
\begin{abstract}
  The simplest transport problem, namely maxflow, is investigated on critical
  percolation clusters in two and three dimensions, using a combination of
  extremal statistics arguments and exact numerical computations, for
  power-law distributed bond strengths of the type $P(\sigma) \sim
  \sigma^{-\alpha}$.  Assuming that only cutting bonds determine the flow, the
  maxflow critical exponent $\ve$ is found to be $\ve(\alpha)=(d-1) \nu +
  1/(1-\alpha)$.  This prediction is confirmed with excellent accuracy using
  large-scale numerical simulation in two and three dimensions.  However, in
  the region of anomalous bond capacity distributions ($0\leq \alpha \leq 1$)
  we demonstrate that, due to cluster-structure fluctuations, it is not the
  cutting bonds but the blobs that set the transport properties of the
  backbone.  This ``blob-dominance'' avoids a cross-over to a regime where
  structural details, the distribution of the number of red or cutting bonds,
  would set the scaling.  The restored scaling exponents however still follow
  the simplistic red bond estimate.  This is argued to be due to the existence
  of a hierarchy of so-called minimum cut-configurations, for which cutting
  bonds form the lowest level, and whose transport properties scale all in the
  same way. We point out the relevance of our findings to other scalar
  transport problems (i.e.  conductivity).
\end{abstract}
\pacs{64.60.Ak,74.25.Fy,73.50.-h}
\maketitle
\section{Introduction}
The transport properties of percolation clusters have been a subject of
interest for many years~\cite{SA94,SN98}. A natural problem to study is e.g.
conductivity, and one often complicates it further by using random bond
``strengths'' $\sigma$ with a power-law tail of the form $ \sigma^{-\alpha}$
~\cite{KSD79,SN82,HFSD85,MGMC86,FHST87,MC88,LTR88,OLN91,SJC01}.  In the first
place, because this allows to represent continuum
percolation~\cite{HFSD85,FHST87} and thus get closer to some actual physical
realizations of percolation.  A second reason why these systems are
interesting is the equivalence~\cite{AHLH71} between transport on strongly
disordered systems and percolative transport~\footnote{ Although this
  equivalence is strict only in the $\alpha\to 1 $ limit, it holds with
  corrections for $1>\alpha$ as well.}. \\
Transport critical exponents on these systems are found to depend on $\alpha$,
which means that strict universality is lost. The original observation that
transport exponents become nonuniversal is due to Kogut and
Straley~\cite{KSD79}, who used mean-field type arguments.  Later
Straley~\cite{SN82}, with the help of the nodes-links-blobs~\cite{\nlb}
picture of the backbone, concluded that the conductivity exponent $t$, such
that $\Sigma \sim (p-p_c)^t$, is the maximum of the universal exponent
$t_0=(d-2)\nu + \zeta$ and the $\alpha$-dependent exponent $\bar t(\alpha)=
(d-2) \nu + 1/(1-\alpha)$. Here $\nu$ is the correlation length exponent and
$\zeta$ measures the contribution of blobs to the resistance between two
points on the backbone, for the case of constant conductances. For the
conductivity problem thus there is a crossover from the universal exponent
$t_0$ for $\alpha<\alpha_c$ to $\bar t(\alpha)$ in the ``anomalous regime''
$\alpha>\alpha_c$.  Although not without some controversy
initially~\cite{BBN81,MGMC86,MC88,LTR88}, this result is by now well
established~\cite{MC88,LTR88,OLN91,SJC01}.

It is somehow surprising that $\bar t(\alpha)$ can be analytically calculated
in the ``anomalous regime'', given that the universal exponent $t_0$, which
applies to the arguably simpler case of constant conductance, has not been
analytically derived up to now. The difficulty in deriving $t_0$ resides in
that $\zeta$ is determined by the blobs, and thus one would require detailed
information~\cite{BHF96} about the internal structure of the blobs.

On the contrary, it has been argued by several
authors~\cite{MGMC86,BHF96,SJC01} that $\bar t(\alpha)$ in the anomalous
regime is determined by the cutting-bonds alone. Since these form linear
chains of typically $L^{1/\nu}$ bonds at $p_c$~\cite{CC82}, the resulting
conductivity exponent is easily derived.  The argument to support the belief
that blobs are irrelevant in the anomalous regime seems to be roughly the
following: an exceedingly small conductivity falling on a blob has little
effect on the overall conductance, because there are many alternative parallel
paths. On the other hand, if this small conductivity is located on a cutting
bond it will certainly dominate the system conductance. While this argument is
true in principle, this reasoning misses the fact that the number of cutting
bonds is itself a fluctuating quantity. The issue of blob-irrelevance has been
considered by Machta and collaborators~\cite{MGMC86} using a hierarchical
model for the backbone, to reach similar conclusions. However, as noted by the
authors, their model does not include structural fluctuations. We will
demonstrate in this work that it is in fact the blobs and not the cutting
bonds, that determine the critical transport properties, even in the anomalous
regime. This as we will see, is due to structural fluctuations. However, the
resulting transport properties turn out to be the same as those given by the
most simplistic red bond estimate. 

A related critical transport problem, which is relevant for disordered
superconductors, is that of determining the critical current density $J_c =
I_c L^{-(d-1)}$ that a percolation network can sustain, and which above $p_c$
behaves as $J_c \sim
(p-p_c)^\ve$~\cite{DRC79,LHSN87,OOAN88,GGWC88,HOAN94,MDHN99}. This problem has
a simple geometrical interpretation. Finding the maximum flow of current, or
\emph{maxflow}, is equivalent to finding the surface across the system, on
which the sum of critical currents of the bonds is maximized. As we will draw
advantage of this analogy later on, we note that this surface is called a
\emph{mincut} in computer science language~\cite{PSC98,CLRI93,RBL89}.

In this paper our goal is to present the first comprehensive study of the
maxflow problem on percolation clusters. This is motivated by the following
observations. First, this is the simplest transport problem that one can think
about, and has not been as such discussed much in the literature.  Second, we
are able to use to our advantage recent developments~\cite{ADME01} on
combinatorial optimization algorithms, in the context of disordered systems.
Here one can use a three-step approach, in which first a critical spanning
cluster is set, its backbone is pruned out, and finally that is used for the
max-flow/min-cut problem. Each stage is solved with one of the powerful graph
optimization algorithms for the particular problem, as discussed later.

In the simplest version of the maxflow problem, all present bonds have the
same critical current, or \emph{capacity} $i_c$ and absent bonds have $i_c=0$.
At criticality, a typical percolating cluster is a linear chain of cutting
bonds and thus $I_c = i_c$. From this observation plus the usual scaling
relation $J_c(L) \sim L^{-\ve/\nu}$, one concludes that~\cite{DRC79}
$\ve=(d-1)\nu$. This result is consistent with experiments~\cite{DRC79,GGWC88}
and numerical simulation~\cite{OOAN88,MDHN99}.  In a more realistic model,
each present bond has a random capacity $i_c$ with power-law distribution
$P(i_c) \sim i_c^{-\alpha}$. This is for example the case for continuum
percolation models~\cite{LHSN87,OOAN88,HOAN94}. A simple extension of the
``typical cutting bond string'' argument gives $v(\alpha)=(d-1)\nu +
1/(1-\alpha)$ as we show later in \Sec{sec:maxflow}. 

In the following we will find it useful to compare the conductivity and
critical current problems to each other. This comparison is done by
interpreting the random bond variables $i_c$ alternatively as bond
conductances $\sigma$ or as bond capacities $i_c$.  Consider for example two
bonds with $\sigma_1$ and $\sigma_2$ connected in parallel. The resulting
conductance \hbox{$\sigma_{par} = \sigma_1 + \sigma_2$} is then the same as
the maximum current $I_{max}$ that can flow if $\sigma_j$ are capacities. If
these bonds are instead connected in series, then
\hbox{$\sigma_{series}=(\sigma_1^{-1}+\sigma_2^{-1})^{-1}$} and
$I_{max}=min(\sigma_1,\sigma_2)$ are no longer equal. However the series
conductance can be written as~\cite{LP89} \hbox{$\sigma_{series} =
  min(\sigma_1,\sigma_2) (1+ \beta)^{-1}$}, with
$\beta=min(\sigma_1,\sigma_2)/max(\sigma_1,\sigma_2)$. In the limit of strong
disorder ($\alpha\to 1$), $\beta$ is typically negligible. We conclude that,
in this limit, also in the series case the conductance equals exactly the
maximum current obtained by interpreting $\sigma_j$ as capacities $i_c$.
Therefore in the $\alpha\to 1$ limit, the resistive current problem and the
superconducting current problem (maxflow) are equivalent, at least for all
structures that can be solved by a combination of series and parallel bond
reductions~\footnote{Similar ideas were put forward in~\cite{MGMC86}, for
  hierarchical structures which are reducible by series and parallel
  transformations.}. Moreover as shown in \Sec{results:results}, we find that
the equivalence noticed above is valid not only in the $\alpha\to 1$ limit but
for a range of $\alpha$ values, for strings of bonds in series.

In deriving $\alpha$-dependent exponents, both for $v(\alpha)$ and $\bar
t(\alpha)$, the assumption is made that the backbone always contains
$L^{1/\nu}$ cutting bonds. While this is true typically, the number of cutting
bonds is in fact a fluctuating variable whose distribution may extend down to
zero in the form of a power law (see later). The existence of such
fluctuations has been noted by some works previously~\cite{KS86b,HAR97}, but
their role in transport properties has not been considered.  These number
fluctuations, we will show in \Sec{sec:maxflow}, do modify the transport
exponent that results from a string of cutting bonds.  Then by analyzing the
conceptually and numerically simple maxflow problem, we will be able to show
that in fact blobs cannot be neglected. The net outcome, which we justify by a
heuristic hierarchical picture, is that although the simplest cutting bond
scaling (without fluctuations) is restored, it is in fact the blobs that set
this scaling behavior.

The structure of the rest of the paper is as follows. Section
\ref{sec:maxflow} presents the analytical discussion, based on a ``fluctuating
number of cutting-bonds'' picture. In Section \ref{numerics} we go through
one-by-one the numerical methods employed, the findings about structural
fluctuations, and some further numerical analysis of the extremal statistics
aspects. Section \ref{sec:blob} contains the results concerning the maxflow
problem, and some details of interest that can be determined from analyzing
large statistics.  Section \ref{concl} finishes the paper with a discussion.
\section{Critical current density}
\label{sec:maxflow}
We consider diluted lattices where the maximum supercurrent $i_c$ that a
present bond can sustain is a random variable distributed between $0$ and $1$
according to
\begin{equation}
P(i_c)= (1-\alpha) c^{-\alpha},
\label{eq:pofi}
\end{equation}
with $\alpha<1$.

Let $I_c$ be the maximum supercurrent (or \emph{maxflow}) that the whole
system, given a set of values $\{ i_c \}$, can sustain. The average current
density $J_c$ is then \hbox{$J_c = <I_c>/L^{(d-1)}$}, and goes to zero at
$p_c$ as
\begin{equation}
J_c \sim (p-p_c)^\ve
\end{equation}
Right at $p_c$, and for a system of finite linear size $L$, usual finite-size
scaling arguments~\cite{SAI94,BHF96} imply that
\begin{equation}
J_c(p_c,L) \sim L^{-\ve/\nu},
\label{eq:vfss}
\end{equation}
where $\nu$ is the percolation correlation length exponent.  The
nodes-links-blobs picture of the percolation cluster~\cite{\nlb} tells us
that, right at $p_c$, there is typically a single connected path through the
sample. This path is a sequence of multiply connected regions (blobs)
connected by strings of singly connected bonds, also called \emph{cutting
  bonds}. The average number of cutting bonds is of order $L^{1/\nu}$ at
$p_c$~\cite{CC82}.

We now start by considering the maximum flow $f^*$ allowed by a string of $n$
cutting bonds, and which obviously equals the least capacity among the $n$
bonds. The typical least value $f^*_n$ among a collection of $n>>1$ random
numbers $i_c$ with probability $P(i_c)$ satisfies
\begin{equation}
\int_{0}^{f^*_n} P(i) di = 1/n
\label{eq:extremal}
\end{equation}
Thus
\begin{equation}
f^*_n = n^{-1/(1-\alpha)}
\label{eq:ityp}
\end{equation} 
On a system of linear size $L$ at $p_c$, the average number of cutting bonds
is $L^{1/\nu}$~\cite{CC82}. In replacing this one obtains $f^*_n \sim
L^{-1/\nu(1-\alpha)}\sim L^{d-1} J_c \sim L^{(d-1) -\ve(\alpha)/\nu}$ and thus
from \Eqn{eq:vfss},
\begin{equation}
\ve(\alpha)=(d-1) \nu + 1/(1-\alpha)
\label{eq:vtrivial}
\end{equation} 
as advanced in the introduction.  

This typical-$n$ argument however neglects the fact that $n$ is a fluctuating
number.  Since ${\mathcal{P}}(n)$ actually has a power law tail extending down
to $n=0$~\cite{HAR97}, this neglect turns out not to be correct for quantities
that depend on $1/n$ as \Eqn{eq:ityp}.

We now present a more careful treatment, which takes into account the
fluctuations in $n$.  It is known~\cite{HAR97} that
\hbox{${\mathcal{P}}_L(n)=(n^*_L)^{-1} \hat {\mathcal{P}}(n/n^*_L)$}, where
$\hat {\mathcal{P}}(\hat n)$ is a size-independent function, and \hbox{$n^*_L
  \sim L^{1/\nu}$}~\cite{CC82}.  Since for the purpose of our discussion all
that matters is the behavior of ${\mathcal{P}}(n)$ as $\hat n \to 0$, we take
for simplicity \hbox{$\hat {\mathcal{P}}(\hat n) = (1+a) \hat n^{a}$}, for
$0 < \hat n \leq 1$. Thus,
\begin{equation}
{\mathcal{P}}(n) =  (1+a) (n^*_L)^{-(1+a)} n^{a},
\label{eq:pofn}
\end{equation} 
for $1\leq n\leq n^*_L$. We will for the moment assume that $n$ cannot be
zero.

Let now $f$ be the minimum among $n$ numbers $x$ distributed with probability
$P(x)$. The distribution $m_n(f)$ of $f$ is determined as
\begin{eqnarray}
\nonumber
m_n(f)&=&nP(f)\left \{ 1- \int_0^f P(x)dx
\right \}^{n-1} \\ &\approx& nP(f) e^{-n \int_0^f P(x)dx}.
\end{eqnarray}
Because of the strong exponential suppression that occurs for $f$ larger
than $f^*_n$ defined by \Eqn{eq:extremal}, we can approximate $m_n(f)$ by
\begin{equation}
m_n(f) \approx  \left \{
\begin{array}{llll}
&n P(f)  \qquad &\hbox{if}& 0<f \leq f^*_n\cr
 \cr
&0  \qquad &\hbox{if}&  f>f^*_n\cr
\end{array}
\right . 
\label{eq:mnapprox}
\end{equation}
Now allowing for the fact that $n$ fluctuates, the probability distribution
function (PDF) of the maxflow $f$ through a string of cutting bonds is
\begin{eqnarray}
\nonumber m(f) &=& \int_1^{\infty} dn {\mathcal{P}}(n) m_n(f) \\
&=&(1+a)(n^*_L)^{-(1+a)}\int_1^{n^*_L} dn n^{a} m_n(f) ,
\label{eq:fpdf}
\end{eqnarray}
for $0<f<1$. 
From \Eqn{eq:ityp} and \Eqn{eq:mnapprox} we conclude that, for a given value
of $f$, the only nonzero contributions in \Eqn{eq:fpdf} come from $n$ values
which are smaller than \hbox{$\eta(f)=f^{-(1-\alpha)}$}. Thus
\begin{eqnarray}
m(f) &=& 
\frac{P(f)(1+a)}{(n^*_L)^{1+a}}\int_1^{\hbox{min}(n^*_L,\eta(f))} 
 n^{a+1} dn 
\label{eq:fpdf2}
\end{eqnarray}
Defining $f_{typ} = (n^*_L)^{-1/(1-\alpha)}$, $\kappa=(a+1)(1-\alpha)$ and
$\lambda=\kappa/(a+2)$, this last expression can be written as
\begin{equation}
m(f) = \left \{
\begin{array}{llll}
&\lambda f_{typ}^{-1} \left (\frac{f}{f_{typ}}  \right )^{-\alpha}
\qquad &\hbox{if}& 0<f<f_{typ}\cr
 \cr
&\lambda f_{typ}^{-1} \left (\frac{f}{f_{typ}}  \right )^{-(1+\kappa)}
\qquad &\hbox{if}& f_{typ}<f<1. \cr
\end{array}
\right . 
\label{eq:fpdf3}
\end{equation}
This gives the PDF for the maxflow $f$ through a string of cutting bonds on a
system of size $L$, allowing for fluctuations in the number $n$ of bonds on
the string. The strength of the fluctuations of $1/n$ is characterized by the
exponent $\kappa$, which in turn depends on $a$. If $a\to \infty$
(non-fluctuating limit) $m(f)$ is nonzero only for $f<f_{typ}$. Thus $<f> \sim
f_{typ}$ and \Eqn{eq:vtrivial} is recovered in this case.  However it is known
that $a\approx 0.22$ in two dimensions~\cite{HAR97}.

For general $a$ and $\alpha$, $m(f)$ has a power-law tail with exponent
$(1+\kappa)$ for $f>>f_{typ}$. The importance of this power-law tail is
evidenced by considering the average flow
\begin{eqnarray}
<f> &=& 
\frac{\lambda}{2-\alpha} f_{typ} +
\frac{\lambda}{1-\kappa} \left ( f_{typ}\right )^{\kappa}
\end{eqnarray}
When $\kappa>1$, \hbox{$<f> \sim f_{typ} \sim (n^*_L)^{-1/(1-\alpha)} \sim
L^{-1/\nu(1-\alpha)}$} and \Eqn{eq:vtrivial} is recovered.  However if
$\alpha>a/(a+1)$ ($\kappa<1$), the power-law tail dominates the
average. In this case \hbox{$<f> ~ f_{typ}^{\kappa} >> f_{typ}$}. Therefore
$<f> \sim L^{-(a+1)/\nu}$, and  \Eqn{eq:vfss} implies that in this case,
\begin{equation}
\ve = (d-1) \nu +  a+1
\label{eq:vmodif}
\end{equation}
The meaning of this is clear. If $\alpha$ is large, typical cases with
$\mathcal{O}(n^*_L)$ cutting bonds will only allow an exceedingly small flux
$f$.  The average flow $<f>$ however will be dominated by the very rare cases
in which $n$ is small and for which $f\sim {\mathcal{O}}(1)>>f_{typ}$. So
finally we conclude that, if we idealize the backbone at $p_c$ as a string of
$n$ cutting bonds, and if $\mathcal{P}(n)$ behaves for small $n$ as $n^a$, one
has that
\begin{equation}
\ve(\alpha) = \left \{
\begin{array}{llllll}
&\frac{1}{1-\alpha}+\nu(d-1)  \qquad &\hbox{if}& \alpha<\frac{a}{1+a}\cr
 \cr
& a + 1 +\nu(d-1)  \qquad &\hbox{if}& \alpha>\frac{a}{1+a}
\end{array}
\right . 
\label{eq:vfinal}
\end{equation}

\section{Numerical Results}
\label{numerics}
\subsection{Algorithms}
In this section we test our analytical derivation of $\ve(\alpha)$ of
\Sec{sec:maxflow} in two and three dimensions on large systems, with the help
of powerful combinatorial algorithms~\cite{ADME01}. Percolation backbones are
first generated by means of a matching algorithm~\cite{MA98,ADME01}, for
square and cubic lattices.  We do this by randomly adding bonds one at a time
until a percolation path is first present. At this point the matching
algorithm identifies the conducting backbone, exactly at the percolation point
for each sample. Alternatively one could fix the density of present bonds to a
value close to the infinite system critical density $p_c$, and then identify
the percolating backbone with the same algorithm. However, our procedure has
the advantage that no separate estimate is necessary for $p_c$.

For each percolating backbone, capacities are drawn from the given
distribution, and the maxflow is calculated by means of a flow augmentation
algorithm (see \cite{ADME01} for a review of the max-flow problem).
The efficiency of the maxflow algorithm is highly
increased when working on the backbone only, so we are able to analyze
thousands of samples for each value of $\alpha$. In this way we estimate
numerically the average flow at $p_c$ for several linear sizes $L$, and from
its scaling properties $\ve(\alpha)$ is derived.  The largest sample sizes
studied were $L=4000$ in two dimensions and $L=120$ in three dimensions.
These are mostly set by the CPU usage of the combination of the matching and
flow algorithms, which in turn is dominated for $L$ large by the scaling of
the matching part. The maxflow code is actually sub-linear in $n=L^2$ in CPU
time, since the mass of the backbone scales with its fractal dimension. Notice
that once the backbone of a sample has been established, it can be used for
several consequent maxflow determinations for different $\alpha$ to save
CPU-time.  In the Appendix we present an idea for an optimal
algorithm for this problem.


\subsection{Results}
\label{results:results}
\begin{figure}[htb]
\centerline{
\psfig{figure=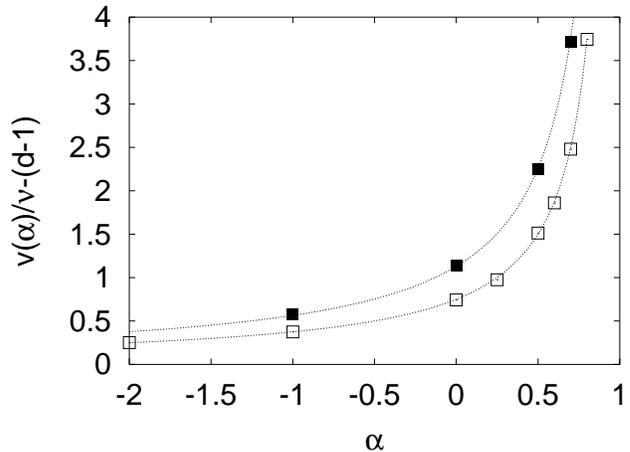,width=9cm,angle=270} }
\caption{{} Maxflow exponents measured for 2d (empty squares) and 3d (filled
  squares) percolation clusters at $p_c$. Dotted lines indicate the
  theoretical result $\ve(\alpha)/\nu - (d-1) = 1/\nu(1-\alpha)$, with
  $1/\nu=0.75$ (2D) and 1.13 (3D).  The maximum linear size simulated was
  $L=4000$ in 2d and $L=120$ in 3d. Error bars are smaller than symbol sizes.
  }
\label{fig:vmeasured} 
\end{figure}

Results are shown in \Fig{fig:vmeasured}.
Our numerical simulation results confirm \Eqn{eq:vtrivial} nicely. However the
saturation of $\ve(\alpha)$ predicted by \Eqn{eq:vfinal} for $\alpha>a/(a+1)$
does not occur. Notice that $a$ is not a universal exponent but depends on the
ensemble.  
\begin{figure}[htb]
\centerline{
\psfig{figure=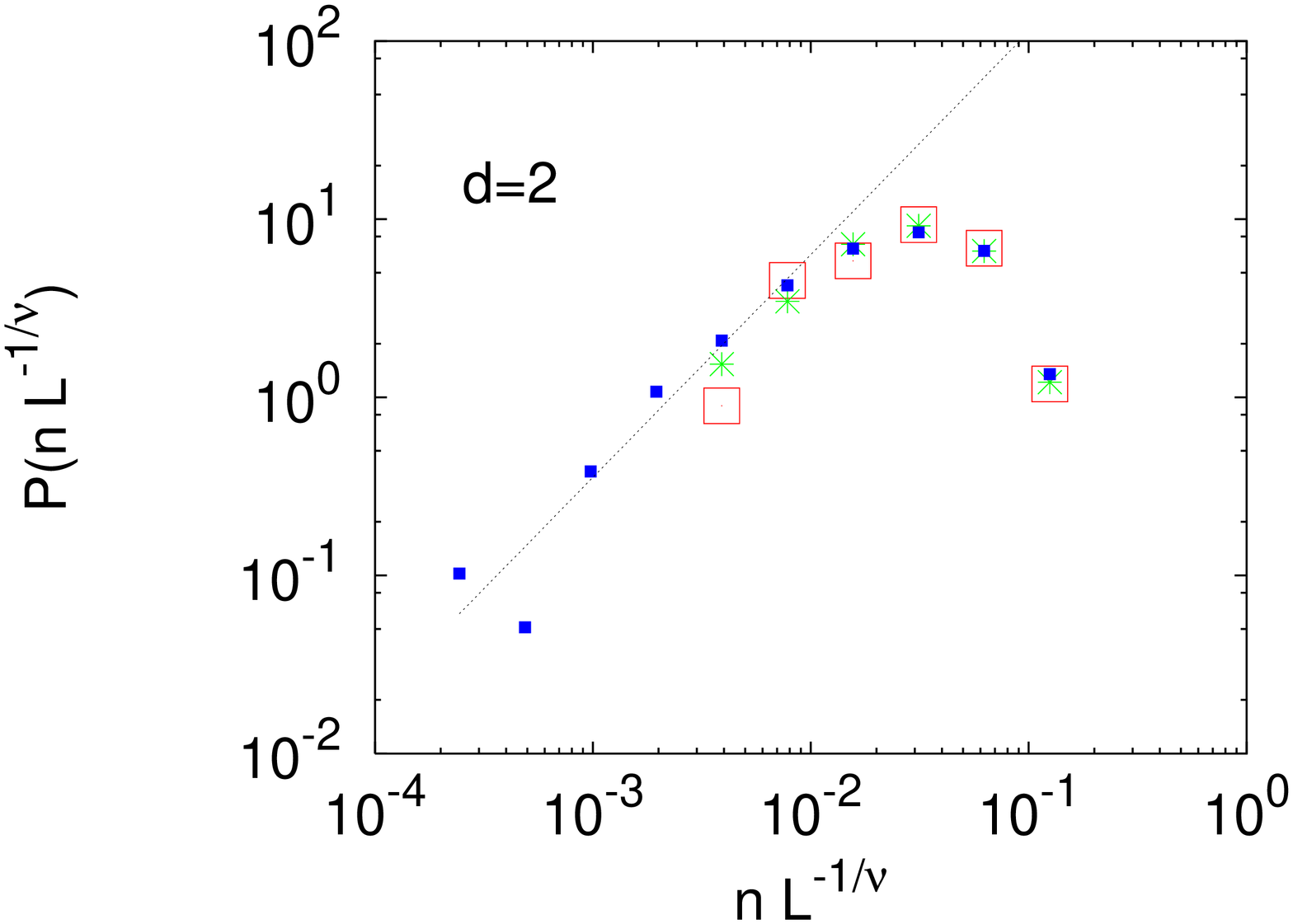,width=9cm,angle=270} 
}\centerline{
\psfig{figure=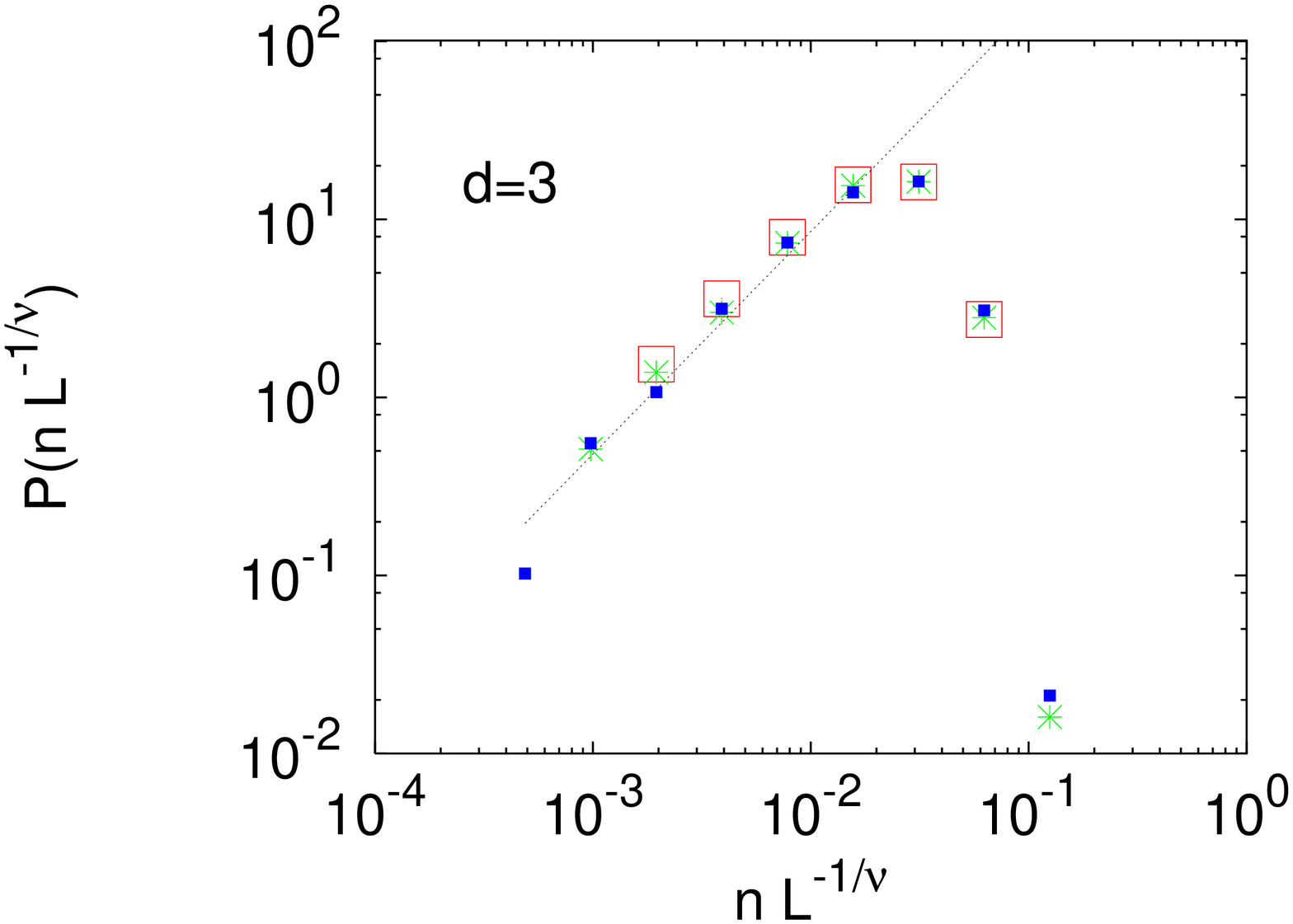,width=9cm,angle=270} }
\caption{{} Distribution of the number $n$ of
  cutting bonds on percolation clusters in two (top: $L=32$ (empty squares),
  $128$ (asterisks) and $1024$ (full squares)) and three (bottom: $L=16$ (empty
  squares), $32$ (asterisks) and $64$ (full squares)) dimensions, in terms of
  the reduced variable $N=n/L^{1/\nu}$.  We find that $P(n)$ is consistent
  with a power-law $n^a$ for small $n$ (dashed lines). Within numerical
  accuracy, the exponent $a=1.25$ both in two and three dimensions.}
\label{fig:pofn} 
\end{figure}
For example if the ensemble is determined by fixing $p=p_c$, numerical
measurements and RG calculations~\cite{HAR97} give $a\approx 0.22$.
Additionally, roughly $20\%$ of the connected samples have zero cutting
bonds~\cite{KS86b,HAR97}, that is, $\mathcal{P}(n) \sim 0.20 \delta(n) + c
n^{0.22}$ for small $n$.

However other ensembles can be considered. Consider for example the
percolation cluster defined by Ambegaokar \emph{et al} construction, in which
conductances are laid down on the lattice in order of increasing conductivity
until a percolating path is created~\cite{AHLH71}. At least the last
conductance to be laid down is a cutting bond, so $\mathcal{P}(0)=0$.
Experimentally this situation is realized when superconductive samples grow
percolatively by deposition~\cite{HOAN94}. In this case the point at which the
supercurrent is nonzero for the first time is defined by the first appearance
of a connected path, not by a fixed density of occupied bonds.

As we add bonds one at a time, our numerical simulations correspond to this
case rather than to fixing $p=p_c$. Our measurements of the distribution of
the number of cutting bonds indicate (\Fig{fig:pofn}) that $\mathcal{P}(n)\sim
n^a$ for small $n$, with $a\approx 1.25$ in two and three dimensions.

Different ensembles give rise to different distributions of the number of
cutting bonds, and specifically to different values of $a$ so, if
\Eqn{eq:vfinal} were to hold for percolation clusters, the resulting transport
exponent would be ensemble-dependent. However, our maxflow measurements on
percolation clusters are consistent with \Eqn{eq:vtrivial} for all $\alpha$,
without any sign of saturation.

In view of the failure of percolation clusters to show the predicted exponent
saturation, we first confirmed the validity of \Eqn{eq:vfinal} for strings of
cutting-bonds. We did so by numerically studying strings of bonds whose number
$n$ is distributed according to (\ref{eq:pofn}), and whose conductances (or
capacities, for the maxflow problem) are distributed according to
(\ref{eq:pofi}). The maximum flow is simply the least critical current $i_c$.
Alternatively, bond capacities $i_c$ may be interpreted as conductances, in
which case the resulting conductance for the whole string is simply
\hbox{$\sigma = 1/\sum_{j=1}^n 1/i_c(j)$}.
\begin{figure}[htb]
\centerline{
\psfig{figure=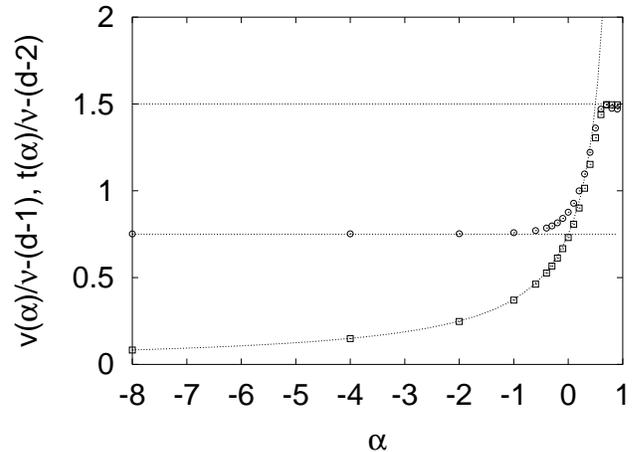,width=9cm,angle=270} 
}
\caption{{} Maxflow (squares) and ohmic current (circles) scaling exponents, as
  numerically estimated for strings of cutting bonds. The number of bonds $n$
  on the string is distributed according to \protect \Eqn{eq:pofn}, with
  $a=1.00$ and $n^*_L=L^{1/\nu}$ with $1/\nu=0.75$. Averages were taken over
  $10^7$ samples, for $L=32, 128, 512,2048$ and $8192$. Notice that (apart
  from a trivial shift) both critical exponents saturate to $(a+1)/\nu$ for
  large $\alpha$. For the maxflow exponent this is the behavior predicted by
  \protect \Eqn{eq:vfinal}. The fact that the (shifted) conductivity exponent
  has the same behavior for all $\alpha>0$ indicates that the sum of
  resistances along the string is dominated by the largest one, in this
  regime. }
\label{fig:cutting.bonds} 
\end{figure}
We find for these strings of cutting bonds (\Fig{fig:cutting.bonds}) that
\Eqn{eq:vfinal} is satisfied very accurately.  \Fig{fig:cutting.bonds} also
shows that the conductivity and maxflow exponents are the same for $\alpha>0$,
indicating that the conductance is dominated by the least $i_c$ value in that
regime. We conclude that \Eqn{eq:vfinal} is exact for strings of
cutting-bonds.  Thus the failure of \Eqn{eq:vfinal} for percolation clusters
simply means that these \emph{do not} behave as strings of cutting bonds do.
In other words, for $\alpha$ near $1$, it is not correct to approximate a
percolation cluster as a string of cutting bonds.
\section{The role of blobs}
\label{sec:blob}
\subsection{Structural fluctuations}
Our results (\Fig{fig:vmeasured}) show that the maxflow exponent $\ve(\alpha)$
follows \Eqn{eq:vtrivial}, although fluctuations in the number $n$ of cutting
bonds, which exist and are relevant in real percolation clusters, were
disregarded in its derivation. So we face a somehow paradoxical situation,
since a naive calculation gives the correct result (\Eqn{eq:vtrivial}), while
a seemingly more careful calculation that takes into account the fluctuations
in $n$ (\Eqn{eq:vfinal}) does not. As mentioned in the previous section, this
means that our assumption that the maximum flow is determined by the cutting
bonds alone needs to be revised. In order to test this assumption, we
separately measure the maximum flow allowed by cutting bonds and by blobs,
which we call $m_c$ and $m_b$ respectively, for each percolation cluster.  The
overall maximum flow is the minimum of these.  The procedure works such that
one picks first the smallest of the cutting bond capacities, and then assigns
to them an infinite capacity. Then the maxflow is found, which is now given by
the minimal blob min-cut (configuration).
\begin{figure}[htb]
  \centerline{{\bf a)}\psfig{figure=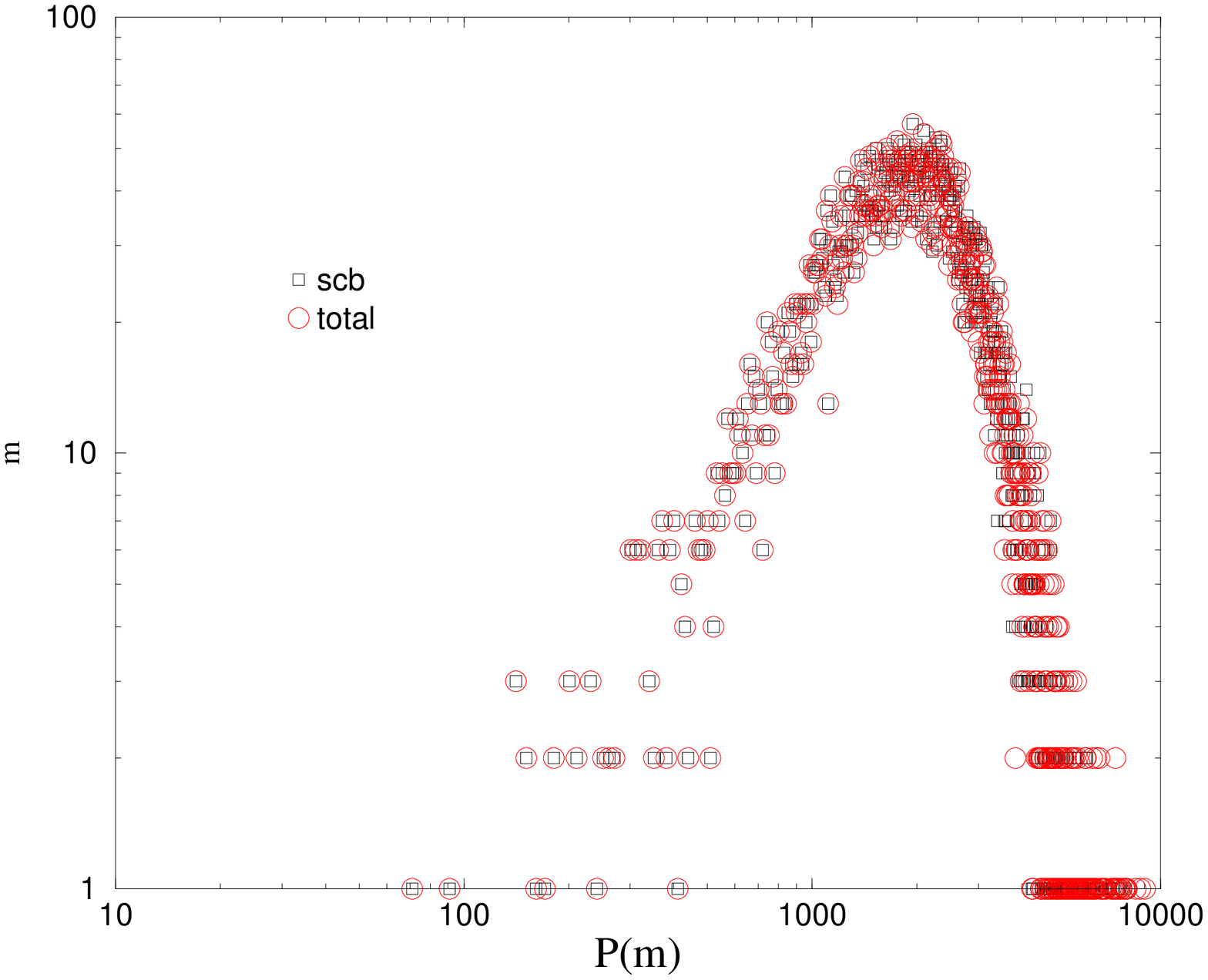,width=7cm,angle=270} }
  \centerline{{\bf b)}\psfig{figure=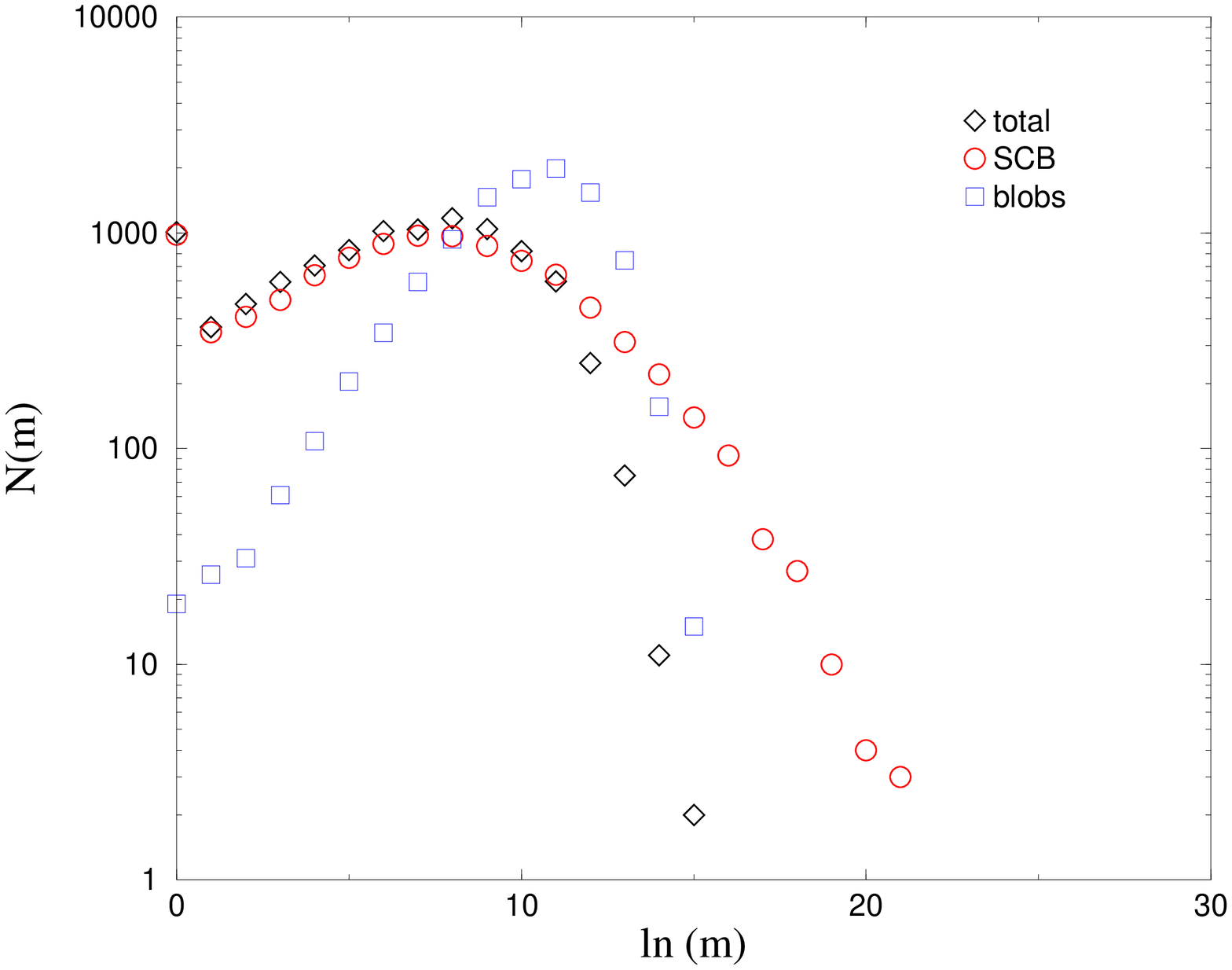,width=7cm,angle=270} }
  \centerline{{\bf c)}\psfig{figure=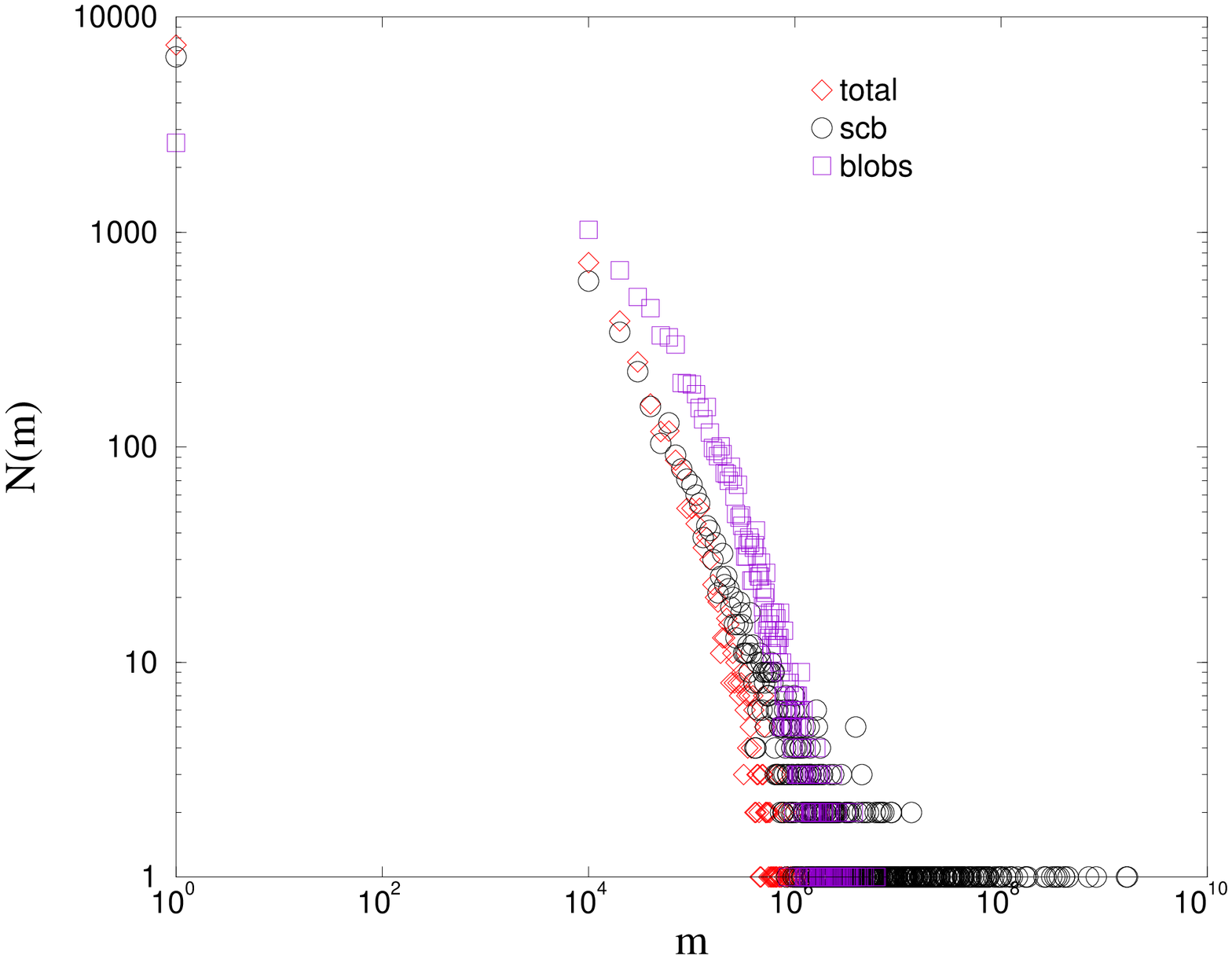,width=7cm,angle=270} }
\caption{{} Probability distribution for the maxflow allowed by cutting bonds
  (squares), blobs (circles) and resulting maxflow (crosses), which is the
  minimum of both. Results are shown for $L=256$ in two dimensions. From top
  to bottom, the disorder exponent is: $\alpha=0.0, 0.5$ and $0.7$.}
\label{fig:blct} 
\end{figure}
\begin{figure}[htb]
\centerline{
\psfig{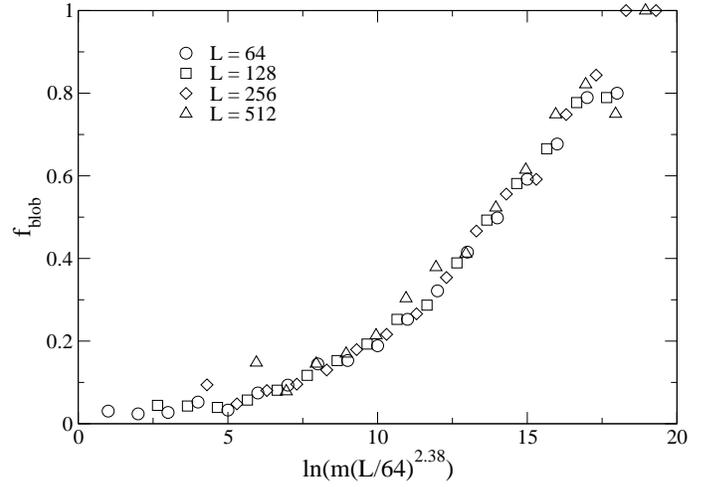} 
}
\caption{{} Probability distributions for the blob-dominated fraction.
The data are collapsed by scaling with the average maxflow. $m=-0.7$.}
\label{fig:fracblct} 
\end{figure}
\begin{figure}[htb]
\centerline{
\psfig{figure=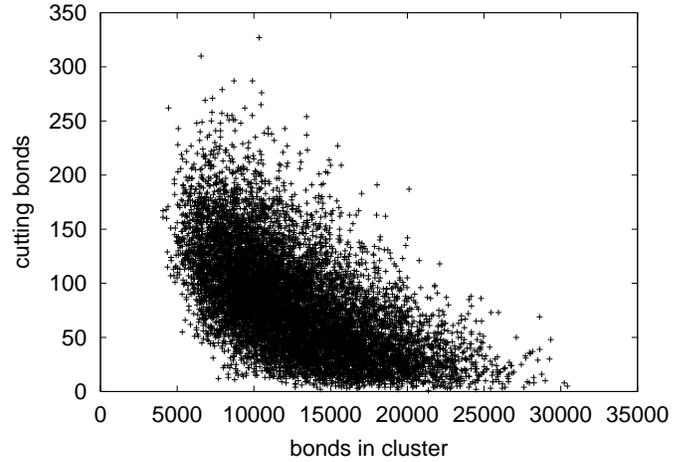,width=9cm,angle=270} 
}
\caption{{} The number of SCB's for $L=256$ vs. the backbone
mass, sample-to-sample. 10000 samples.}
\label{fig:scbvsnbonds} 
\end{figure}
\begin{figure}[htb]
\centerline{
\psfig{figure=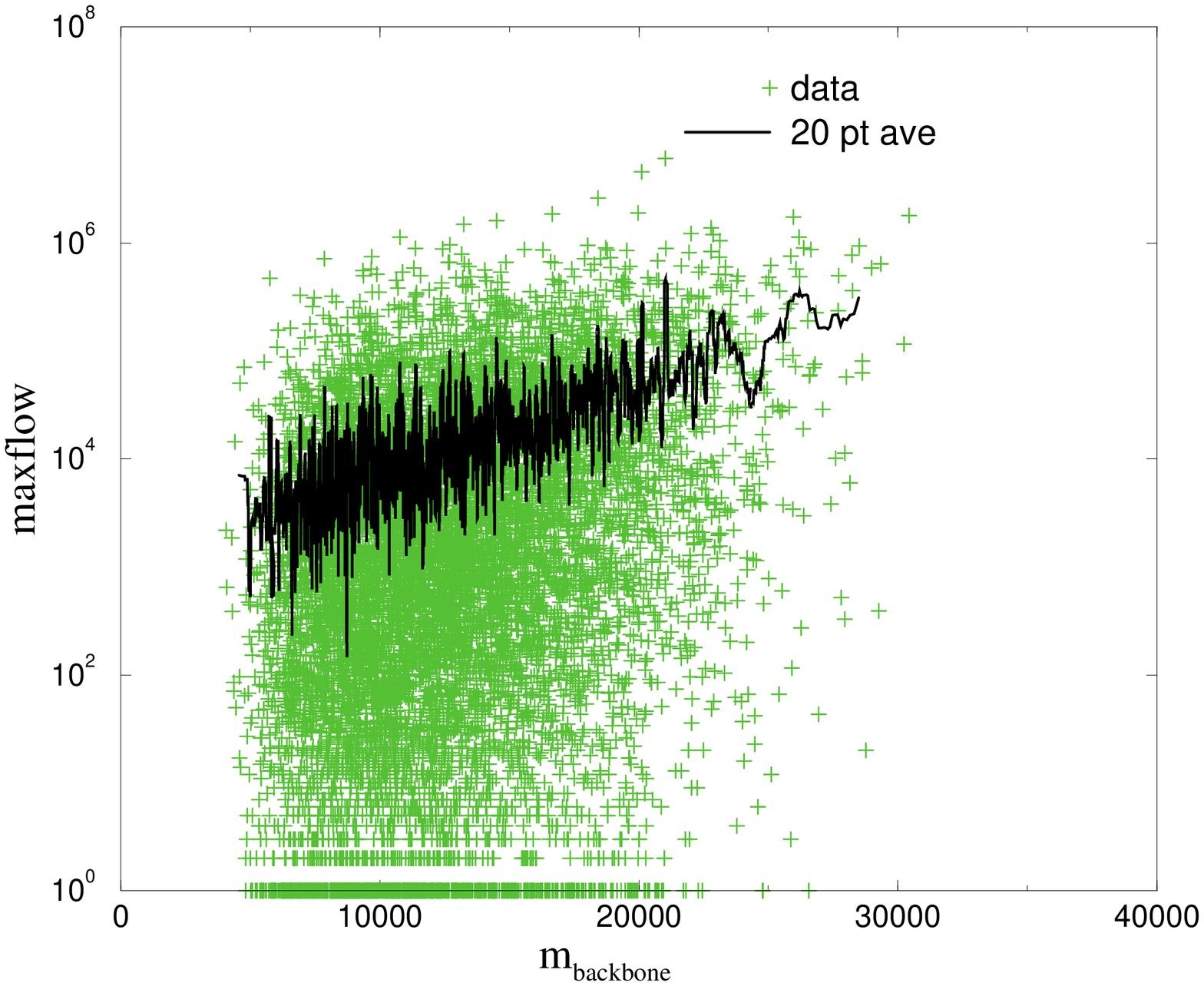,width=9cm,angle=270} 
}
\caption{{} The maxflow, for $L=256$ and $m=-0.7$, vs. the actual
backbone mass. The average is a running average over 20 samples,
with consecutive masses from 10$^4$ samples.}
\label{fig:mfvsnbonds} 
\end{figure}
Figures \ref{fig:blct} a to c show how the PDF of $m_c$ (cutting-bond flow)
and the total PDF vary with $\alpha$. For non-anomalous values $\alpha=0$
(Fig.~\ref{fig:blct}a), the distribution is centered around a well-defined
mean value.  With  increasing $\alpha$ one enters the anomalous regime,
and the PDF develops a power-law tail. This would be expected to result from
the cutting bonds, while the blob flows $m_b$ have a much narrower
distribution, decaying roughly exponentially for large flows.  This means
that, when $m_c$ is large, most probably $m_b$ will be much smaller and thus
the overall flow will be determined by $m_b$.  Thus, although our derivation
of \Eqn{eq:vfinal} is correct for strings of cutting bonds, it is the blobs
that determine the flow in those rare cases in which $m_c$ is large.
Therefore the power-law tail in $P(m_c)$, which is responsible for the
saturation of $\ve(\alpha)$ at large values of $\alpha$ in \Eqn{eq:vfinal}, is
suppressed by blobs on percolation clusters. Figure \ref{fig:fracblct}
illustrates this by showing that the fraction of cases - for a given maxflow
$m$ - that are dominated by the blob contribution follow a separate PDF. The
collapse is not completely perfect, since there may be a very slight trend in
the total fraction of blob-dominated cases with increasing $L$. On the other
hand, the variances of the maxflow distributions scale as expected (as the
mean). It is worth mentioning that the distribution of $k$-cuts (number of
bonds in the min-cut) is roughly exponential, so that $\langle k \rangle$ is
of the order of 1.4 $\dots$ 1.5 for $\alpha=0.7$.

It is also worth pointing out that there are cross-correlations between the
structural quantities on one hand, and between the structure and the maxflow
on the other hand. These are illustrated in Figures \ref{fig:scbvsnbonds} and
\ref{fig:mfvsnbonds}. In a system with a given $L$ it is after a moment's
deliberation rather clear that there may be an inverse correlation between the
\emph{ number} of cutting bonds and the sample-to-sample weight of the
backbone. We have not tried to measure this relationship quantitatively, but
given such a relation it is no surprise (Fig.~\ref{fig:mfvsnbonds}) that the
mass of the backbone correlates strongly with the maxflow value.
\subsection{Blob dominance}
\begin{figure}[htb]
\centerline{
\psfig{figure=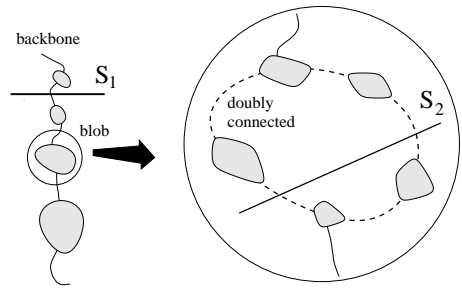,width=6cm,angle=0} 
}
\caption{{} Backbone structure. }
\label{fig:backbone} 
\end{figure}

In order to prove that our hypothesis, namely that the blobs set the maxflow
scale, is correct, we still have to show that the blob flow $m_b$ has the
right scaling properties, i.e.  $\overline{m}_b \sim L^{-1/\nu(1-\alpha)}$. A
complete calculation of the maximum flow allowed by blobs would require
detailed information about the blob's internal structure.  However an estimate
can be obtained from the following arguments. It is known that the backbone at
$p_c$ has a hierarchical, or self-similar, structure~\cite{CC82}. At the top
level of this hierarchy, the backbone itself can be thought of as a string of
singly connected (cutting) bonds interspersed with blobs. Blobs in turn are
loops made of doubly connected bonds interspersed with smaller blobs and so
on, as depicted in \Fig{fig:backbone}.  This hierarchical structure has its
counterpart in a similar classification of surfaces which separate the
backbone into two pieces (cuts). At the top level of this hierarchy are the
surfaces $\{S_1$\} that cut the backbone at just one bond, next come those
surfaces $\{S_2\}$ that cut the backbone at exactly two bonds, etc. The
capacity $C(S)$ of a cut $S$ is defined as the sum of the capacities $i_c$ of
the bonds crossed by it.  Because of the maxflow-mincut theorem, the maximum
flow equals the minimum of the cuts' capacities.  Our assumption that cutting
bonds alone determine the maximum flow is equivalent to minimizing the
capacities among the $S_1$ alone. We now describe how the next level $S_2$ in
this hierarchy can be analyzed.
Coniglio~\cite{CC82} has shown that the derivative of the spanning probability
$p'$ with respect to $p$ is proportional to the average number of cutting
bonds $<n>$. An extension of his reasoning, due to Kantor~\cite{KS86b}, allows
one to write the second derivative of $p'(p)$ with respect to $p$ at $p_c$ as
\hbox{$\partial^2 p'/\partial p^2|_{p_c} \sim < n(n-1)-2N_2>|_{p_c}$}, where
$n$ is the number of cutting bonds and $N_2$ is the number of pairs of doubly
connected bonds.  Because by definition $\partial^2 p'/\partial p^2=0$ at
$p_c$~\footnote{Here we depart slightly from Kantor's reasoning, who justifies
  this last point as due to duality. Duality is not necessary and therefore
  \hbox{$2<N_2>=<n^2>-<n>$} for any lattice and dimension, at $p_c$.}, one
finds that \hbox{$2<N_2>=<n^2>-<n>$} at $p_c$.  Since $<n^k> \sim
L^{k/\nu}$~\footnote{ This results from the fact that the distribution of $n$
  decays exponentially fast for large $n$. See \protect \Fig{fig:pofn} and
  ref.~\protect\cite{HAR97}. }, we conclude that the typical number of pairs
of doubly connected bonds at $p_c$ is $N_2 \sim L^{2/\nu}$. However this alone
is not enough to estimate the typical maximum flow allowed by doubly connected
bonds, for they might be grouped into blobs in different ways. Fortunately the
total number $n_2$ of doubly connected bonds at $p_c$ can also be
calculated~\cite{CC82}, and it turns out to be $<n_2> \sim L^{1/\nu}$. This
means that the blob statistics is dominated by one large ring of roughly
$L^{1/\nu}$ bonds and therefore containing a number of pairs of doubly
connected bonds which is of order $L^{2/\nu}$. Using this information we can
now estimate the maximum flow allowed by blobs at the level of doubly
connected bonds. This large blob dominates the maximum flow since lesser
blobs, located somewhere else along the backbone, will allow a larger flow.
Thus one has to find the maximum flow for two parallel strings, each
containing $L^{1/\nu}$ cutting bonds. The typical flow allowed by each string
is of order $L^{-1/\nu(1-\alpha)}$ and therefore the typical maximum flow
allowed by doubly-connected blobs, which is twice this, is of the right order.

Our reasoning for doubly connected bonds only considers typical cases, i.e.
fluctuations in the number of doubly connected bonds are disregarded. If a
particular cluster, in addition to having a small number of cutting bonds,
also has a small number of doubly connected bonds, then the next levels in
this hierarchy would be relevant.  The same sort of reasoning can be used at
all levels in the hierarchy of cuts, but because the algebra becomes too
complicate for triply connected bonds already, we did not test this in detail.
However it seems safe to assume that $min\{C(S_k)\} \sim L^{-1/\nu(1-\alpha)}$
for all $k>2$ as well. Additionally notice that, in order for the mincut to be
located at triply connected bonds, it is necessary that the numbers of singly
and doubly connected bonds be simultaneously small, an occurrence which
arguably has a  small probability.

We then see that, in those rare cases in which the number of singly connected
bonds is small (they allow a large flow), blobs take their role thus limiting
the flow to a value which is typically of order $L^{-1/\nu(1-\alpha)}$. This
then shows that the correct value of the exponent $\ve(\alpha)$ is given by
\Eqn{eq:vtrivial}. Similar behavior is of course expected for other transport
properties, e.g. conductivity, in the limit of anomalous distributions of bond
strengths.  This is so because in this limit the resistance of the whole
cluster is dominated by that of the mincut, where conductivities are
interpreted as critical currents.
\section{Conclusions}
\label{concl}
In this article we have demonstrated that the transport problem on percolation
clusters still holds surprises. Our findings deny the widespread notion that,
in the limit of anomalous strength distributions, it is the cutting-bonds
alone that determine the transport properties.  We show analytically, and
confirm numerically that, if blobs could be neglected (because of their
allowing a larger maxflow than cutting bonds) then the overall system's
behavior would be strongly dependent on the ensemble (the cutting-bond PDF
tail exponent). This ensemble-dependence would come about because the number
of cutting bonds has a ``broad'' distribution extending down to zero. However
the predicted ensemble-dependence is not there, as we show numerically on
large two and three-dimensional systems.  Using scaling arguments we then
demonstrate that it is in fact the blobs that finally determine the average
maxflow.  However, we are forced to finish with the paradoxical conclusion
that though the expected mechanism for the maxflow, namely cutting-bond
dominance, does not work in the anomalous regime (large $\alpha$), the
original cutting-bond estimate for the transport exponent is nevertheless
restored by the limiting effect of the blobs.

\acknowledgments This collaboration was partially supported by V\"ais\"al\"an
Foundation, Finland. CFM acknowledges finantial support by CONACYT, M\'exico,
through research project 36256-E.  CFM also wishes to acknowledge the kind
hospitality of the Laboratory of Physics, HUT, where parts of this work were
done.  MJA is supported by the Academy of Finland Center of Excellence
program.
\appendix{\bf Appendix - an optimal algorithm:}
We note that the augmenting path -method is better here than in the general
maxflow problem so-called push/relabel preflow algorithms~\cite{GTA88} enjoy
the most popularity).  This is since the structure of the backbone is
essentially one-dimensional, the number of augmentations remains small, of the
order of one.  To remind the reader, such an algorithm consists of \emph{ flow
  augmentations}, which are repeated till the min-cut is formed (by a surface
of blocked bonds) and maxflow is reached.  For each augmentation one needs to
establish a path from the ``source'' to the ``sink'', which can be done
e.~g.~. by using shortest-distance path methods~\cite{ADME01}.

The one-dimensional nature means that the backbone can be decomposed into
strings of subsequent cutting bonds $C_i$ and and blobs separating such
strings $B_i$. Thus the structure is equivalent to the one-dimensional series
$\dots C_i B_i C_{i+1}\dots$.  In principle one may thus write a more
efficient algorithm by abandoning the lattice structure, and describing the
internal geometry of each $B_i$ separately. Thus an optimal version of the
algorithm would entail the following steps: \emph{ i):} establish the
structure ($B_i$, $C_i$). \emph{ ii):} Find an augmenting path along the
chain, across all $B_i$.  \emph{ iii):} Augment flow, that is: find the
smallest capacity in the $C_i$, and the smallest capacity in all the $B_i$.
This is $f_1$.  \emph{ iv):} if $f_1$ equals the minimal cutting bond capacity
stop, otherwise augment (subtract $f_1$ from $C_i$, and the paths inside
$B_i$). \emph{ v):} update the paths inside those $B_j$, only, where a bond
was saturated by $f_i$ ($i=1$ to begin with). Goto \emph{ iii)}. We have used,
instead, an Euclidean background for the maxflow part, since the scaling of
the matching program is indeed the bottleneck.
\bibliographystyle{prsty} 
\bibliography{Books,Moukarzel,PHASETRAN,maxflow,}
\end{document}